\documentclass[conference]{IEEEtran}
\IEEEoverridecommandlockouts
\usepackage{cite}
\usepackage{amsmath,amssymb,amsfonts}
\usepackage{hyperref}
\usepackage[capitalize]{cleveref}
\usepackage{algorithmic}
\usepackage{graphicx}
\usepackage{textcomp}
\usepackage{xcolor}
\def\BibTeX{{\rm B\kern-.05em{\sc i\kern-.025em b}\kern-.08em
    T\kern-.1667em\lower.7ex\hbox{E}\kern-.125emX}}

\usepackage{bm}
\usepackage{mathtools}

\usepackage{xcolor}

\newcommand{\est}[1]{\hat{#1}}
\renewcommand{\vec}[1]{\bm{\mathrm{#1}}}
\newcommand{\obs}[1]{\tilde{#1}}
\newcommand{\set}[1]{\mathcal{#1}}
\newcommand{\freq}[1]{\bar{#1}}

\newcommand{\gaussian}{\mathcal N}
\newcommand{\R}{\mathbb{R}}

\newcommand{\ep}{\varepsilon}
\newcommand{\Si}{\Sigma}
\newcommand{\si}{\sigma}
\newcommand{\al}{\alpha}
\newcommand{\la}{\lambda}
\newcommand{\La}{\Lambda}

\begin{document}

\title{A Bayesian Perspective on Uncertainty Quantification for Estimated Graph Signals
\thanks{
We acknowledge partial funding by the Ministry of Culture and Science (MKW) of the German State of North Rhine-Westphalia (``NRW Rückkehrprogramm'') and the European Union (ERC, HIGH-HOPeS, 101039827). Views and opinions expressed are however those of the author(s) only and do not necessarily reflect those of the European Union or the European Research Council Executive Agency. Neither the European Union nor the granting authority can be held responsible for them.
}
}

\author{\IEEEauthorblockN{Lennard Rompelberg}
\IEEEauthorblockA{\textit{Department of Computer Science} \\
\textit{RWTH Aachen University}\\
Aachen, Germany \\
rompelberg@cs.rwth-aachen.de}
\and
\IEEEauthorblockN{Michael T. Schaub}
\IEEEauthorblockA{\textit{Department of Computer Science} \\
\textit{RWTH Aachen University}\\
Aachen, Germany \\
schaub@cs.rwth-aachen.de}
}

\maketitle

\begin{abstract}
We present a Bayesian perspective on quantifying the uncertainty of graph signals estimated or reconstructed from imperfect observations.
We show that many conventional methods of graph signal estimation, reconstruction and imputation, can be reinterpreted as finding the mean of a posterior Gaussian distribution, with a covariance matrix shaped by the graph structure.
In this perspective, assumptions of signal smoothness as well as bandlimitedness are naturally expressible as the choice of certain prior distributions;
noisy, noise-free or partial observations are expressible in terms of certain likelihood models.
In addition to providing a point estimate, as most standard estimation strategies do, our probabilistic framework enables us to characterize the shape of the estimated signal distribution around the estimate in terms of the posterior covariance matrix.

\end{abstract}


\section{Introduction}

When observing signals on graphs, the observations often do not perfectly capture the true signal.
On the one hand, observations are often noisy and thus subject to error~\cite{shuman2013,chen2015}.
On the other hand, we may only have access to observations on a subset of nodes~\cite{anis2014,gadde2015,tsitsvero2016}.
Thus, subsequent tasks often require some form of error correction and imputation of missing parts.
Notably, the resulting estimated or reconstructed signals are associated with uncertainty about how close they are to the true signal.

Some earlier work focused on designing sample subsets of nodes that enable perfect reconstruction from partial observations, assuming that the signal is bandlimited and observations on the sample subset are noise-free~\cite{anis2014,tsitsvero2016}.
In constrast, noisy observations over the entire graph may be corrected by regularization that incentivizes some notion of smoothness~\cite{shuman2013}.
More recent work aims to find sample sets, which given specific assumptions about the signal space, minimize uncertainty about the reconstructed signal~\cite{tanaka2020,gadde2015,bai2020}.
This uncertainty is usually otherwise left unexamined.
However, with the exception of models that allow perfect reconstruction, some uncertainty persists even after such error correction, in particular whenever observations are noisy.
Furthermore, this uncertainty is not generally a global or node-wise scalar, but is instead structured based on the graph and sampling strategy.
Indeed, we will show that uncertainty about the signal mean is not improved over the observational noise level, even though the mean is often of key interest.

In this article, we utilize that current approaches in signal estimation, reconstruction, and imputation can be understood in terms of Bayesian inference.
We will show that smoothness and bandlimit assumptions can be encoded as prior distributions over the signal space.
Likewise, both noisy and partial observations can be encoded as likelihood distributions.
The associated posterior distribution then provides a description of the uncertainty about this reconstructed signal.
Processing this distribution, instead of a point estimate, in subsequent tasks enables us to gauge the trustworthiness of results.

In the remainder of this article, we first lay out the required foundations of graph signal processing in~\cref{sec:gsp}, as well as mention some preliminaries and related work in~\cref{sec:related-work}.
We then present the framework for probabilistic graph signal estimation via Bayesian inference in~\cref{sec:bayesian-inference}, and show that it generalizes smoothness and subspace assumptions, complete and partial observations, and noisy and noise-free observations.
We also leverage the probabilistic model to quantify uncertainty depending on direction, and highlight the benefits of this approach.
Finally, we showcase the ability to quantify uncertainty for an example in~\cref{sec:experiments}.

\subsection{Graph Signal Processing}
\label{sec:gsp}

We consider undirected graphs $\set{G} = (\set{V}, \set{E})$, where $\set{E} \subseteq \binom{\set{V}}{2}$.
A graph signal is a function $x\colon \set{V} \to \R$ that assigns a real number to each node.
However, we will fix an enumeration of the nodes $\{v_1, \dots, v_n\} = \set{V}$ and write graph signals as vectors $\vec{x} = [x_1, \dots, x_n]^\top \in \R^n$, where $x_i \coloneqq x(v_i)$.
Throughout this article, we will denote observed signals as $\obs{\vec{x}}$, estimated signals as $\est{\vec{x}}$, and true signals \emph{or} free variables as $\vec{x}$.
Given the adjacency matrix $A \in \R^{n \times n}$ with $a_{ij} = 1$ if $\{i,j\} \in \set{E}$ and $a_{ij} = 0$ otherwise, and the degree matrix $D \coloneqq \operatorname{diag}(A \vec{1})$, we define the \emph{graph Laplacian} as $L \coloneqq D - A$.

In analogy to the Fourier transform on continuous signal spaces, we define the graph Fourier transform (GFT) through the graph Laplacian:
Fixing a spectral decomposition $L = V \La V^\top$ with $V = [\vec{v}_1, \dots, \vec{v}_n]$ and $\La = \operatorname{diag}(\vec{\la})$, the eigenbasis $V^\top$ serves as the Fourier transformation.
The Fourier transform of a signal is $\freq{\vec{x}} \coloneqq V^\top \vec{x}$, where $\freq{x}_i$ encodes the coefficient by which the eigenvector $\vec{v}_i$ occurs in the signal $\vec{x}$.
We quantify the \emph{quadratic variation} of a graph signal $\vec{x}$ as
\begin{equation}
    \operatorname{TV}_2(\vec{x}) \coloneqq \vec{x}^\top L \vec{x} = \sum_{i=1}^{n} \la_i \freq{x}_i^2,
\end{equation}
i.e., the occurrences of components associated with high eigenvalues contribute more towards quadratic variation~\cite{marques2020}.

\subsection{Preliminaries and Related Work}
\label{sec:related-work}

Let us briefly mention existing approaches for signal estimation and reconstruction.
Some methods yield optimization-based point estimates, others do make use of Bayesian inference, in particular to minimize variance-based metrics.

\subsubsection{Laplacian Quadratic Regularization}
Let $\obs{\vec{x}} \in \R^n$ be an observation over the entire graph.
We then derive an estimation
\begin{equation}
    \label{eq:penalized-optimization}
    \est{\vec{x}} = \arg\min_{\vec{x} \in \R^n} \big(\|\obs{\vec{x}} - \vec{x}\|^2 + \al \vec{x}^\top L \vec{x}\big)
\end{equation}
for some $\alpha \geq 0$, with solution $\est{\vec{x}} = (I + \al L)^{-1} \obs{\vec{x}}$, i.e., we assume closeness to the observed signal, but also assume smoothness, so we penalize quadratic variation~\cite{shuman2013}.

\subsubsection{Perfect Reconstruction of Bandlimited Signals}
\label{sec:perfect-reconstruction-intro}
Let $\vec{x}$ be in the subspace $\set{U} \coloneqq \operatorname{span}(U)$ for some $U \in \R^{n \times n_g}$.
If $U$ is the subset of columns of the Laplacian eigenbasis $V$ associated with eigenvalues $\la \leq b$, then $\vec{x}$ is $b$-bandlimited.
Let $\obs{\vec{x}} = S^\top \vec{x}$ be the observed, sampled signal for some sampling matrix $S \in \R^{n \times n_s}$, where we denote the subspace $\set{S} \coloneqq \operatorname{span}(S)$.
Often, $S$ is a rearrangement of column vectors of the identity matrix $I$, but this need not be the case.
Graph sampling commonly also considers a sampling set $\set{V}_s$ with $|\set{V}_s| = n_s$ for some fixed sampling budget $n_s$ - then, $S$ consists of the columns of $I$ corresponding to $\set{V}_s$~\cite{tanaka2020}.
Then if $\set{U}$ and $\set{S}^\bot$ have no intersection except at the origin and jointly span $\R^n$, we can perfectly reconstruct the true signal as $\vec{x} = U (S^\top U)^{-1} \obs{\vec{x}}$;
otherwise, $\est{\vec{x}} = U (S^\top  U)^+ \obs{\vec{x}}$ is an optimal approximation~\cite{tanaka2020}.

\subsubsection{Minimum Variance Sampling}
In graph sampling under smoothness assumptions, we often aim to find a sampling set of a given size $n_s$ which minimizes some covariance-based metric~\cite{tanaka2020,tanaka2020a,bai2020,gadde2015,sakiyama2019,chamon2018}.
These approaches either explicitly or implicitly employ a Bayesian framework to obtain the covariance matrix of a Gaussian posterior distribution.
Subsequently, the sample set is chosen such that some metric based on the covariance matrix, e.g., trace, determinant, or maximum eigenvalue, is minimized~\cite{tanaka2020}.
In this article, we want to focus on the covariance matrix itself, instead of aggregate functions thereof, to show that uncertainty may differ significantly depending on direction in the signal space $\R^n$, and should thus be considered in subsequent processing.

\subsubsection{Gaussian Processes on Graphs}
Finally, Borovitskiy et al. apply Gaussian Processes to signals on graphs~\cite{borovitskiy2021}.
Due to the discrete nature of graphs, Gaussian processes on graphs are equivalent to Gaussian distributions~\cite{rasmussen2006}.
Thus, their approach recovers a posterior Gaussian distribution through Bayesian inference.
However, they only derive priors that are uncommon in graph signal estimation and reconstruction.
Moreover, their work does not focus on the uncertainty structure described by the posterior covariance matrix.
We show that a similar method can be derived for the more common smoothness and subspace priors, considering both noisy and partial observations.

\section{Bayesian Inference of Graph Signals}
\label{sec:bayesian-inference}

We now leverage Bayesian inference to obtain a posterior distribution over graph signals, given a noisy observation and an improper prior distribution that arises from the graph structure.
Let $L$ be the Laplacian of a fixed graph.
Assume that the \emph{true} signal $\vec{x}$ is drawn from a prior Gaussian distribution $\gaussian(\vec{0}, L_\ep^{-1})$, where $L_\ep = L + \ep I$ for some $\ep > 0$ to ensure invertibility.
This assumption is equivalent, and indeed often characterized as the signal being generated by a Gaussian Markov random field (GMRF).
However, as we subsequently generalize to other priors, we will characterize it as a Gaussian distribution.
The corresponding probability density function (PDF) is
\begin{equation}
    p(\vec{x}) = (2\pi)^{-\frac{n}{2}} \det(L_\ep)^{\frac{1}{2}} e^{-\frac{1}{2} \vec{x}^\top L_\ep \vec{x}}.
\end{equation}
Intuitively, signals with high quadratic variation are approximately assigned low probability densities.
We further assume that the \emph{observation} $\obs{\vec{x}}$ is subject to i.i.d. additive Gaussian noise, i.e., is drawn from a Gaussian likelihood distribution $\gaussian(\vec{x}, \si^2 I)$ centered on $\vec{x}$.

\begin{figure}
    \includegraphics[width=\linewidth]{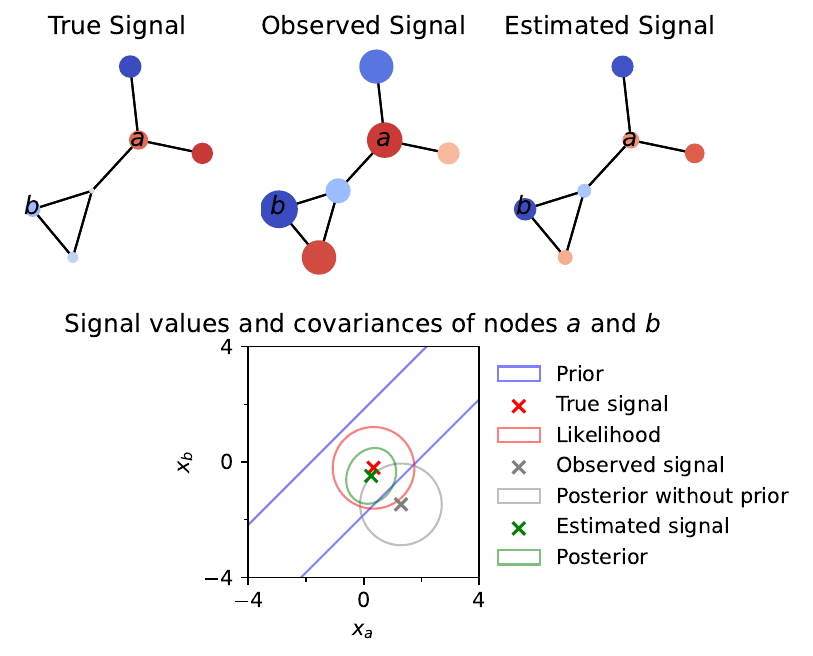}
    \caption{
        Top: A true signal $\vec{x} \sim \gaussian(\vec{0}, L_\ep^{-1})$, observed signal $\obs{\vec{x}} \sim \gaussian(\vec{x}, \si^2 I)$, and estimated signal $\est{\vec{x}} = (I + \si^2 L)^{-1} \obs{\vec{x}}$ for $\ep = 10^{-6}$ and $\si^2 = 1$.
        Bottom:
        The signal values for nodes $a$ and $b$ as points in $\R^2$, with ellipses at the unit variance contour line of the corresponding distributions.
    }
    \label{fig:pairwise-covariances}
\end{figure}

As both likelihood and prior are Gaussian, the posterior is also Gaussian with closed formulas for mean and covariance~\cite{murphy2007}.
In our case, we obtain
\begin{equation}
    \label{eq:posterior-epsilon}
    \vec{\mu}_\ep = (I + \si^2 L_\ep)^{-1} \obs{\vec{x}}
    ,\ 
    \Si_\ep = (\si^{-2} I + L_\ep)^{-1}
\end{equation}
defining posterior distribution of the graph signal.
Importantly, the posterior distribution remains a well-defined Gaussian distribution even in the limit $\ep \to 0$:
\begin{equation}
    \label{eq:posterior-limit}
    \begin{aligned}
        \vec{\mu} = \lim_{\ep \to 0} \vec{\mu}_\ep &= (I + \si^2 L)^{-1} \obs{\vec{x}} \\
        \Si = \lim_{\ep \to 0} \Si_\ep &= (\si^{-2} I + L)^{-1}.
    \end{aligned}
\end{equation}
The involved distributions are visualized in~\cref{fig:pairwise-covariances}.

We now show that the mean of this posterior distribution is exactly the estimate returned by the penalized optimization problem in~\eqref{eq:penalized-optimization} for $\al = \si^2$.
Thus, the covariance matrix of the posterior distribution characterizes the uncertainty structure of this estimate.
Indeed,~\eqref{eq:penalized-optimization} minimizes the negative log of the posterior PDF:
\begin{equation}
    \label{eq:bayesian-density}
    \begin{aligned}
        p(\vec{x} \mid \obs{\vec{x}})
        &\propto p(\obs{\vec{x}} \mid \vec{x})\ p(\vec{x}) \\
        &\propto e^{- \si^{-2} \|\vec{x} - \obs{\vec{x}} \|^2}\ e^{- \vec{x}^\top L \vec{x}} \\
        \implies
        - \log p(\vec{x} \mid \obs{\vec{x}})
        &\propto \|\vec{x} - \obs{\vec{x}}\|^2 + \si^2 \vec{x}^\top L \vec{x}.
    \end{aligned}
\end{equation}

Signal estimation with a quadratic variation penalty therefore returns the mean of a posterior Gaussian distribution.
Thus, we know that uncertainty about the estimated signal is captured by the covariance matrix $\Si = (\si^{-2} I + L)^{-1}$.

We note that this covariance matrix can be interpreted as assigning scalar values of uncertainty to each eigenvector in the basis $V$ of the graph Laplacian $L$:
First, consider that
\begin{equation}
    \Si = (\si^{-2} I + L)^{-1} = V (\si^{-2} I + \La)^{-1} V^\top.
\end{equation}
Each eigenvector $\vec{v}_i$ of $L$ is then associated with uncertainty $(\si^{-2} + \la_i)^{-1}$.
In particular, the all-ones eigenvector is associated with uncertainty $\si^2$, i.e., the prior provides no improvement upon the noise variance.
More generally, \emph{any arbitrary} direction in signal space, represented by a normalized vector $\vec{z} \in \R^n$ with $\|\vec{z}\|=1$, is associated with uncertainty
\begin{equation}
    \vec{z}^\top \Si \vec{z} = \freq{\vec{z}}^\top (\si^{-2} I + \La)^{-1} \freq{\vec{z}} = \sum_{i=1}^{n} (\si^{-2} + \la_i)^{-1} \freq{z}_i^2.
\end{equation}

\subsection{Partial Observations and Graph Sampling}
\label{sec:partial-observations}

Let us now consider that observations may only be available on a subset of nodes.
As in~\cref{sec:related-work}, we assume that sampling is performed by some matrix $S \in \R^{n \times n_s}$, but require that $S$ has full column rank.
Assume that observations are noisy, i.e., $\obs{\vec{x}} \sim \gaussian(S^\top \vec{x}, \si^2 I)$.
We now cannot obtain the posterior covariance $\Si$ like in~\eqref{eq:posterior-epsilon}, as $L \in \R^{n \times n}$ and $\si^2 I \in \R^{n_s \times n_s}$ now have different dimensions.

Let us therefore consider the distribution of $\obs{\vec{x}}' \in \R^n$ with $S^\top \obs{\vec{x}}' = \obs{\vec{x}}$.
Suppose that $\obs{\vec{x}}' \sim \gaussian\big(\vec{x}, \si^2 (SS^\top)_\ep^{-1}\big)$, where $(SS^\top)_\ep = (SS^\top + \ep I)$ for some $\ep > 0$.
In the limit $\ep \to 0$, this implies that $\obs{\vec{x}} \sim \gaussian(S^\top \vec{x}, \si^2 I)$:
\begin{equation}
    \begin{aligned}
        - \log p(\obs{\vec{x}}' \mid\vec{x})
        &\propto \|S^\top (\obs{\vec{x}}' - \vec{x}) \|^2 \\
        &= \| \obs{\vec{x}} - S^\top \vec{x} \|^2
        \propto - \log p(\obs{\vec{x}} \mid \vec{x}).
    \end{aligned}
\end{equation}
As a result, we can compute the posterior covariance matrix as $\Si_\ep = \big( \si^{-2} (SS^\top)_\ep + L \big)^{-1}$.

In contrast to~\eqref{eq:posterior-limit}, $( \si^{-2} SS^\top + L)$ is not guaranteed to be invertible.
Conventionally, this is characterized as the optimization problem
\begin{equation}
    \label{eq:penalized-optimization-sampled}
    \est{\vec{x}} = \arg\min_{\vec{x} \in \R^n} \big( \si^{-2}\| S^\top\vec{x} - \obs{\vec{x}}\|^2 + \vec{x}^\top L \vec{x}\big)
\end{equation}
having no unique solution.
The probabilistic perspective, however, reveals that $\Si = \lim_{\ep \to 0} (\si^{-2} (SS^\top)_\ep + L)^{-1}$ has eigenvectors whose associated eigenvalue approaches infinity, i.e., directions associated with infinite uncertainty.
However, even though there is no unique solution, we still obtain information about directions \emph{not} associated with infinite variance.
That being said, a simple way to ensure invertibility is to retain $L_\ep$, i.e., define $\Si = (\si^{-2} SS^\top + L_\ep)^{-1}$, which leads to an optimization problem with added penalty
\begin{equation}
    \label{eq:penalized-optimization-sampled-unique}
    \est{\vec{x}} = \arg\min_{\vec{x} \in \R^n} \big( \si^{-2}\| S^\top\vec{x} - \obs{\vec{x}}\|^2 + \vec{x}^\top L \vec{x} + \ep \|\vec{x}\|^2\big).
\end{equation}

\subsubsection*{Noise-Free Observations}

In the literature, it is often assumed that the available observations perfectly capture the true signal value at that node~\cite{anis2014,tsitsvero2016}.
Such noise-free observations may be modeled by $\si^2 \to 0$, so the posterior covariance becomes 
\begin{equation}
    \Si = \lim_{\si^2 \to 0} \lim_{\ep \to 0} (\si^{-2} SS^\top + L_\ep)^{-1}.
\end{equation}
Equivalently, the optimization problem in~\eqref{eq:penalized-optimization-sampled} changes such that any deviation $\|S^\top \vec{x} - \obs{\vec{x}}\|^2$ is infinitely penalized, introducing a hard constraint
\begin{equation}
    \est{\vec{x}} = \arg\min_{\vec{x} \in \R^n} \vec{x}^\top L \vec{x} \quad \text{s.t.}\ S^\top \vec{x} = \obs{\vec{x}}.
\end{equation}

\subsection{Subspace Priors}
\label{sec:subspace-priors}

Let us now assume that the true signal lies in a subspace of $\R^n$.
Specifically, we assume a bandlimit model, where the subspace is $\operatorname{span}(V_b)$ for some bandlimit $b$ (see~\cref{sec:perfect-reconstruction-intro}).
In constrast to the previous smoothness assumption, any coefficient by which an eigenvector component in $V_b$ occurs is equally likely, so the distribution within $\operatorname{span}(V_b)$ is implicitly uniform.
We can recover a subspace assumption as the limit of a Gaussian prior, akin to the case of partial noise-free observations:
\begin{equation}
    \Si_\text{prior} = \lim_{\si^2 \to 0} \lim_{\ep \to 0}
    \si^{2} (I - V_b V_b^\top)_\ep^{-1},
\end{equation}
where $(I - V_b V_b^\top)_\ep = (1+\ep)I - V_b V_b^\top$ for some $\ep > 0$ and $\lim_{\si^2 \to 0} \lim_{\ep \to 0} \si^2 \ep^{-1} = \infty$.
The variance of directions outside the subspace goes to zero, and the variance of directions inside the subspace go to infinity.
In contrast, while the limit case recovers an ideal subspace assumption, the use of non-zero variance $\si^2$ and $\ep$ allows for more relaxed assumptions that are arbitrarily close to a subspace assumption.

\subsubsection*{Perfect Reconstruction}

Let us show that with the probabilistic approach, we can characterize perfect reconstruction through a vanishing covariance matrix, i.e., an uncertainty of zero in every direction.
Consider that we have a subspace prior and perfect observations.
Then the posterior covariance matrix is
\begin{equation}
    \Si = \lim_{\si^2 \to 0} \lim_{\ep \to 0}
    \si^2 \big((SS^\top)_\ep + (I - V_b V_b^\top)_\ep\big)^{-1}.
\end{equation}
Then, if $(SS^\top + I - V_b V_b^\top)$ is invertible, $\Si$ vanishes for $\si^2 \to 0$, i.e., there is absolute certainty in the solution.
Thus, we can characterize perfect reconstructibility as the limit case where the posterior probability distribution concentrates onto a single point.

The limit case places ideal constraints on the signal which make the problem solvable via closed formula, which, however, necessarily features expensive and potentially prohibitive matrix inversion.
Through the derivation of these conditions as limits of Gaussian distributions, we obtain an informed way of solving such problems through optimization:
If we do not take the limits of $\big(\si_{\text{obs}}^{-2} (SS^\top)_\ep + \si_{\text{prior}}^{-2} (I - V_b V_b^\top)_\ep\big)^{-1}$, we obtain a well-defined Gaussian distribution which yields a convex optimization problem with a unique solution and non-zero gradient everywhere else, which approximately preserves the uncertainty structure.

\section{Experiments}
\label{sec:experiments}

\begin{figure}
    \includegraphics[width=\linewidth]{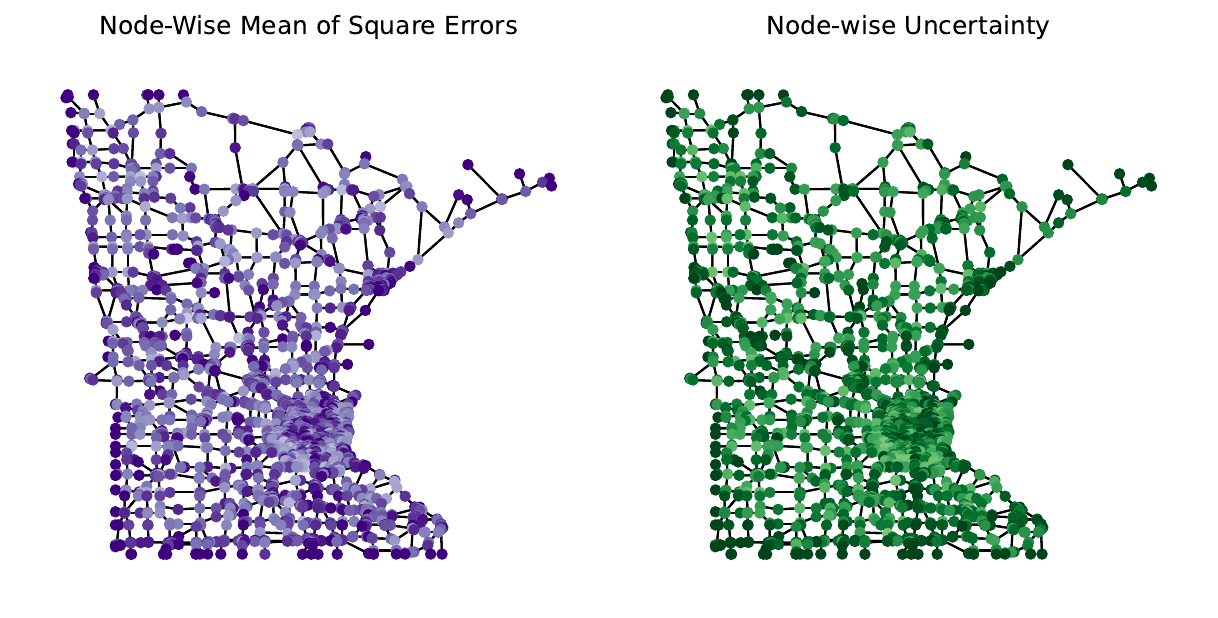}
    \caption{
        Left: The mean over 100 node-wise squared differences between true and estimated signals on the Minnesota traffic network~\cite{rossi2015}.
        Each true signal $\vec{x}$ is drawn independently from $\gaussian(\vec{0}, L_\ep^{-1})$ with $\ep = 10^{-6}$ and independently observed, i.e., $\obs{\vec{x}}$ is drawn from $\gaussian(\vec{x}, \si^2 I)$ with $\si^2 = 3$.
        The estimated signal is computed as $\est{\vec{x}} = (I + \si^2 L_\ep)^{-1}$.
        Right: The node-wise uncertainty represented by the diagonal entries of the posterior covariance matrix.
    }
    \label{fig:squared-error-uncertainty}
\end{figure}

Let us showcase the usefulness of the posterior covariance matrix in quantifying uncertainty based on a direction in signal space.
In this section, we generate synthetic signals over the Minnesota traffic network~\cite{rossi2015} from a Laplacian smoothness prior, and for each true signal, simulate its complete, but noisy observation.
We then compute an estimate from the observation as the mean of the resulting posterior distribution.
We show that the uncertainty in directions associated with nodes of the graph varies significantly between individual nodes.

Note that the uncertainty in a given direction models the expected square error between the estimated and true signal, projected onto that direction.
As an example, we characterize the uncertainty associated with each node $i$ of the graph by considering the associated direction $\vec{e}_i$, the $i$th column of the identity matrix $I$.
Then the uncertainty associated with node $i$ is given by $\vec{e}_i^\top \Si \vec{e}_i$, i.e., the $i$th entry on the diagonal of $\Si$.
In~\cref{fig:squared-error-uncertainty}, we plot the node-wise uncertainty, next to the node-wise mean square error over 100 cases of signal generation, observation, and estimation.

Note that node-wise uncertainty is not a full description of the uncertainty structure, as the off-diagonal entries of the covariance matrix, and thus covariances between pairs of nodes, are ignored.
In fact, because $\Si = (\si^{-2} I + L_\ep )^{-1}$ (see~\cref{sec:bayesian-inference}), non-zero covariance between nodes exists precisely because of the graph structure.

\begin{figure}
    \includegraphics[width=\linewidth]{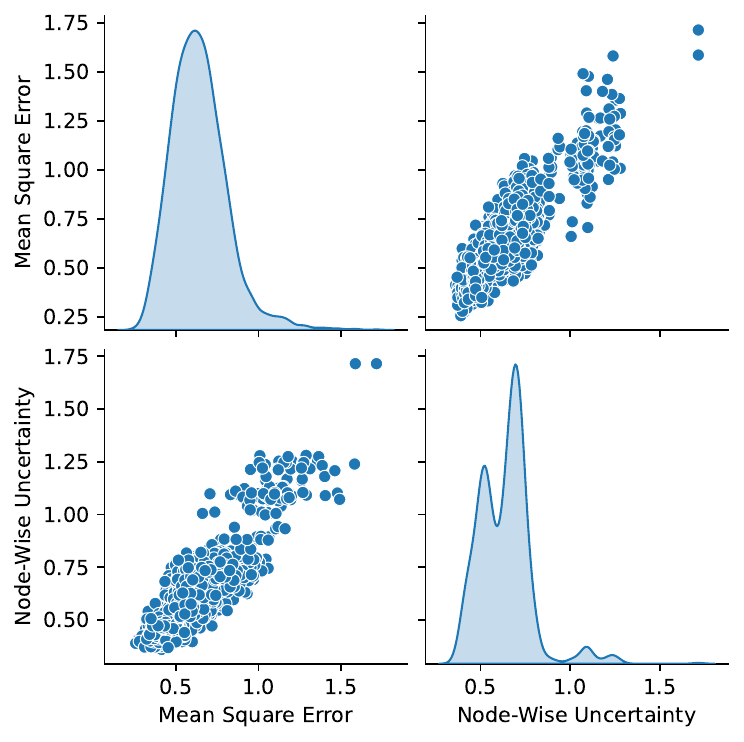}
    \caption{
        Upper left: The distribution of node-wise mean square error over all nodes (see~\cref{fig:squared-error-uncertainty}).
        Lower right: The distribution of node-wise uncertainty over all nodes.
        Upper right and lower left: The relation between node-wise uncertainty and mean square error.
    }
    \label{fig:square-error-variance-relation}
\end{figure}

\Cref{fig:square-error-variance-relation} shows the relation between node-wise uncertainty and node-wise mean square error.
Node-wise uncertainty and node-wise mean square error tend to be similar for each node, so our approach of quantifying uncertainty works for modeling mean square error in the chosen direction in signal space.
On the other hand, uncertainty varies significantly between different nodes, i.e., directions in signal space.

\section{Conclusion}

We presented a framework of generalizing common assumptions about graph signal spaces and observations through Bayesian inference of Gaussian distributions.
We highlighted the use of obtaining entire distributions instead of point estimates as being able to characterize the associated uncertainty structure via the covariance matrix.
As a result, we can make claims about uncertainty depending on direction in the signal space.
Uncertainty may differ significantly depending on direction, so considering the uncertainty structure in subsequent tasks is important to avoid overconfidence in downstream predictions.
Furthermore, we are able to recover information about the signal within any direction associated with finite certainty, even if no unique solution should exist.

\bibliographystyle{IEEEtran}
\bibliography{IEEEabrv,../literature}

\end{document}